\documentclass[
]{ceurart}

\usepackage{listings}
\begin{document}

\copyrightyear{2025}

\conference{LASI Spain 25: Learning Analytics Summer Institute Spain 2025,
  May 26--27, 2025, Vitoria-Gasteiz, Spain}

\title{MOSAIC-F: A Framework for Enhancing Students’ Oral Presentation Skills through Personalized Feedback}


\author[1]{Alvaro Becerra}[%
orcid=0009-0003-7793-2682,
email=alvaro.becerra@uam.es,
]
\address[1]{GHIA Group, School of Engineering, Universidad Autónoma de Madrid, Spain}

\address[2]{BiDA-Lab Group, School of Engineering, Universidad Autónoma de Madrid, Spain}

\author[1]{Daniel Andres}[%
]
\fnmark[1]

\author[1]{Pablo Villegas}[%
]
\fnmark[1]

\author[2]{Roberto Daza}[%
orcid=0009-0005-2109-7782,
email=roberto.daza@uam.es
]

\author[1]{Ruth Cobos}[%
orcid=0000-0002-3411-3009,
email=ruth.cobos@uam.es
]

\fntext[1]{These authors contributed equally.}

\begin{abstract}
In this article, we present a novel multimodal feedback framework called MOSAIC-F, an acronym for a data-driven Framework that integrates Multimodal Learning Analytics (MMLA), Observations, Sensors, Artificial Intelligence (AI), and Collaborative assessments for generating personalized feedback on student learning activities. This framework consists of four key steps. First, peers and professors’ assessments are conducted through standardized rubrics (that include both quantitative and qualitative evaluations). Second, multimodal data are collected during learning activities, including video recordings, audio capture, gaze tracking, physiological signals (heart rate, motion data), and behavioral interactions. Third, personalized feedback is generated using AI, synthesizing human-based evaluations and data-based multimodal insights such as posture, speech patterns, stress levels, and cognitive load, among others. Finally, students review their own performance through video recordings and engage in self-assessment and feedback visualization, comparing their own evaluations with peers and professors’ assessments, class averages, and AI-generated recommendations. By combining human-based and data-based evaluation techniques, this framework enables more accurate, personalized and actionable feedback. We tested MOSAIC-F in the context of improving oral presentation skills.
\end{abstract}


\maketitle

\section{Introduction}
Feedback is a cornerstone of effective learning, serving as a vital tool for students to develop and refine their skills. When delivered thoughtfully, feedback transcends mere correction, guiding students toward deeper understanding, enhanced self-regulation, and sustained progress \cite{hattie2007power}. Feedback should not be viewed as a one-way transmission of information but as an interactive process that fosters reflection, decision-making, and the development of personal improvement strategies \cite{askew2000feedback}.

However, improving students’ skills requires structured and effective feedback, which traditional methods often fail to provide adequately \cite{henderson2019challenges}. Common challenges include subjectivity in evaluation, where feedback can be inconsistent and overly dependent on personal perspectives rather than on clear, standardised criteria, making it difficult for students to understand exactly what aspects of their performance need improvement. Additionally, a lack of specificity and clarity in feedback comments often results in students receiving vague or generic statements that fail to provide actionable guidance, as they do not explicitly highlight both strengths and weaknesses while offering concrete suggestions on how to enhance their performance.

Another major issue is the absence of opportunities for self-reflection and the application of feedback to future tasks. Many students receive feedback at the end of an assignment or course, with no structured mechanism to integrate it into their learning process. As highlighted in \cite{tailab2020use}, students who reviewed video recordings of their performances became more aware of their strengths and weaknesses, allowing them to identify specific areas for improvement.

In order to address these challenges, we introduce MOSAIC-F, an acronym for a data-driven framework that uses Multimodal Learning Analytics, Observations, Sensors, Artificial Intelligence (AI), and Collaborative assessments to generate personalized feedback on student learning activities. In more detail, in this article, we focus on using MOSAIC-F to improve students’ oral presentation skills.

This framework integrates multiple approaches to enhance the assessment process for students’ skills combining peer and self-assessment through standardized rubrics that incorporate both quantitative and qualitative evaluations; sensors and Multimodal Learning Analytics (MMLA) \cite{giannakos2022multimodal} to collect data from physiological signals and audio and camera recordings, enabling more accurate and data-based feedback; and AI to address scalability concerns, analyze multimodal data and synthesize evaluations into comprehensive feedback.

To assess the effectiveness of the MOSAIC-F framework, a case study was conducted for enhancing oral presentation skills with final-year students from the Telecommunication Technology and Service Engineering program at Universidad Autónoma de Madrid (UAM). The study focused on evaluating how multimodal data, peer and professor collaborative assessment, and AI-generated feedback could contribute to the development of professional communication competencies in engineering students.

The study involved 46 students who were required to deliver a 10-minute oral presentation followed by a 5-minute question period as part of their course assessment. Participation was voluntary, and all students signed informed consent forms detailing the data collection process, including video/audio recordings and the use of wearable sensors. Furthermore, ethical considerations were strictly observed, and all collected data were fully anonymized at all times to ensure the protection of participants’ privacy.

The remainder of the article is organized as follows: Section \ref{s:related_works} reviews related work in the areas of Multimodal Learning Analytics and Artificial Intelligence for automated feedback. Section \ref{s:mosaic-f} introduces the MOSAIC-F framework and describes its application in a case study focused on enhancing students’ oral presentation skills. Section \ref{s:multimodal_analyses} outlines the multimodal data analyses that will be carried out as part of the case study. Finally, Section \ref{s:conclusions} presents the conclusions and discusses directions for future work.

\section{Related Works}\label{s:related_works}
\subsection{Multimodal Learning Analytics}
Multimodal Learning Analytics (MMLA) has emerged as a promising field that seeks to capture, integrate, and analyze diverse data sources to better understand learning processes and improve educational outcomes \cite{giannakos2022multimodal}. Unlike traditional Learning Analytics \cite{lang2017handbook}, which often relies solely on log data from learning management systems, MMLA incorporates rich and heterogeneous modalities such as video, audio, physiological signals, gaze data, and behavioral traces.

MMLA has proven to be highly effective in online learning environments, where platforms based on biometrics and behavioral analysis have emerged \cite{baro2018integration,daza2023edbb}, benefiting from recent advances in machine learning and digital behavior understanding. For example, \cite{becerra2023m2lads, becerra2025m2lads, becerra2023user} presents M2LADS, a web-based system that integrates and visualizes multimodal data from MOOC learning sessions. The system collects and synchronizes biometric signals, such as EEG data, heart rate, and visual attention, with behavioral logs and learning performance indicators, offering instructors a dashboard that provides a comprehensive view of learners’ cognitive and emotional engagement during the session. In \cite{becerra2024biometrics}, the authors investigate how biometric and behavioral signals can be used to detect distractions related to mobile phone use during online learning sessions, specifically by analyzing head pose deviations captured in the IMPROVE database \cite{daza2024improve}. Additionally, in \cite{navarro2024vaad} focuses on the analysis of visual attention through eye-tracking data to estimate the specific task the learner is performing.

MMLA has also been successfully applied in face-to-face learning environments. For instance, \cite{spikol2018supervised} presents a multimodal system embedded in physical computing worktables, which captures students’ hand movements, gaze direction, use of programming interfaces, and audio levels to evaluate collaboration and predict the quality of student-generated artifacts. By applying supervised machine learning techniques, the system was able to identify strong performance predictors, such as physical proximity between students and hand motion dynamics. 

Similarly, \cite{garcia2024exploring} explored the use of MMLA techniques to analyze face-to-face teaching practices based on classroom audio recordings. Their system combines deep learning for speaker diarization and machine learning for practice classification to identify instructional methods such as lectures or group work.

\subsection{Artificial Intelligence and Automatic Feedback}
Artificial Intelligence has emerged as a key tool for processing and analyzing multimodal educational data, enabling the detection of hidden patterns, the handling of large data volumes, and the delivery of adaptive feedback \cite{bosch2014s,ekin2025artificial}. More recently, the rise of generative AI has enhanced the capabilities of automated feedback systems—ranging from the generation of written comments to real-time dashboards and alert mechanisms that monitor students’ learning progress \cite{kim2024llm}.

In the case of oral presentations, several studies have focused on developing dashboards that use visualizations to provide students with feedback during practice sessions before delivering their presentations. For instance, in \cite{schneider2017presentation}, the authors introduced a multimodal feedback system that provides real-time analysis of nonverbal communication, including voice volume, posture, gestures, and pauses. The system uses sensor-based technologies such as depth cameras and microphones to capture and interpret students’ nonverbal behavior during a presentation. Based on interviews with public speaking experts, the tool identifies a set of effective and ineffective nonverbal practices, offering automated feedback that aligns with commonly accepted teaching methods. Importantly, the study emphasizes that such systems should prioritize raising students’ awareness and encouraging reflection, rather than enforcing rigid performance standards.

Building on this line of work, in \cite{di2025presentable} an AI-driven tool that supports both the content creation and rehearsal phases of oral presentations was developed and it offers guidance on structuring messages, provides audio and video-based feedback on delivery, and uses self-reflection prompts to foster awareness and improvement.

Similarly, \cite{ochoa2018rap} proposed a low-cost solution that leverages basic sensors, such as webcams and ambient microphones, to analyze key features like gaze direction, posture, voice volume, filled pauses, and visual slide content. Their system generates post-session feedback reports, integrating video and audio recordings to help students identify areas for improvement.

More recently, \cite{ochoa2024openopaf} presented an open-source multimodal system designed to provide automated feedback on oral presentation skills in real time. The system evaluates body language, voice volume, articulation speed, gaze direction, filled pauses, and visual slide design, offering presenters immediate and actionable information. Unlike many prior tools, this system emphasizes accessibility, modularity, and scalability, aiming to support widespread adoption in educational settings and enable further research into automated feedback mechanisms.

Additionally, virtual reality has also been explored in this context \cite{daza2025smartevr, yokoyama2021vr}. For example, \cite{yokoyama2021vr} proposed a VR-based rehearsal system that automatically evaluates students’ presentations using machine learning techniques applied to gesture and movement data captured in immersive environments, providing feedback and enabling self-review through avatar playback.

\subsection{Evaluating Human Versus AI-Based Feedback}

One of the critical aspects in the implementation of automated feedback is how users perceive its credibility, usefulness, and fairness. Prior research comparing human tutors with intelligent tutoring systems has shown that automated feedback can be nearly as effective as human-generated feedback in certain contexts \cite{vanlehn2011relative}. However, important differences emerge in terms of acceptance: students often perceive human-generated feedback as more trustworthy and contextually sensitive \cite{zhang2019co}. \cite{rudian2025feedback} further supports these concerns, showing that students perceived auto-generated feedback as less valuable and emotionally disconnected, especially when they were aware that it was produced by a language model.

Additionally, in \cite{nazaretsky2024ai} a large-scale study was conducted with over 450 university students to assess how the perceived identity of a feedback provider (human or AI) affects student evaluations. They found that students often rated AI-generated feedback lower once its source was revealed, especially in terms of genuineness and credibility, even when the content quality was comparable. While students may recognize the benefits of AI, concerns about transparency, fairness, and overreliance can significantly influence their perception, an insight particularly relevant when evaluating how AI-generated feedback is received \cite{nazaretsky2025critical}.

Despite these concerns, recent studies suggest that generative AI tools like ChatGPT can produce feedback that is both detailed and effective. In \cite{steiss2024comparing}, the authors compared human and AI-generated feedback on student essays. While human assessors provided more nuanced and personalized comments, ChatGPT consistently delivered rubric-aligned feedback that met formal assessment criteria with high precision. Similarly, \cite{wan2024exploring} examined students’ responses to AI-generated comments on conceptual physics questions. Students rated the AI feedback as accurate and, in many cases, more useful than human-generated feedback due to its comprehensiveness and coverage.

At the same time, generative AI offers clear advantages in terms of scalability and immediacy. Unlike human-generated feedback, which can be time-consuming and inconsistent, AI systems can produce instant draft comments, which instructors can then refine to enhance pedagogical value \cite{kloos2024can}.

Given these developments, recent work emphasizes the need to go beyond system performance and address the human factors surrounding educational technology adoption. In particular, involving educators and learners in the design of AI-based feedback systems has shown promise in enhancing both their usability and acceptance. Co-design approaches facilitate the alignment of data-driven tools with classroom needs, build trust among stakeholders, and promote a stronger sense of ownership and engagement in the use of educational technology \cite{ogata2024co}. Along these lines, \cite{topali2023unlock} emphasizes pedagogical grounded feedback interventions based on student data, contextual awareness, and personalization, all of which are critical for increasing students’ trust and acceptance of automated feedback systems.

\section{MOSAIC-F Applied to Enhancing Students’ Oral Presentation Skills}\label{s:mosaic-f}
In this article, we introduce MOSAIC-F, a data-driven framework that integrates Multimodal Learning Analytics, Observations, Sensors, Artificial Intelligence (AI), and Collaborative assessments for generating personalized Feedback. To illustrate the implementation and potential of MOSAIC-F, we present its application in a concrete case study focused on improving the oral presentation skills of students. In this case study, MOSAIC-F involves both professors and students in the feedback process through the following roles:
\begin{itemize}
\item \textbf{Evaluators:} At least one professor and two students evaluate the presenter’s performance.
\item \textbf{Presenter:} The student who delivers the oral presentation.
\item \textbf{Observers:} One student in the audience is monitored using an eye-tracking device, while a research assistant simultaneously annotates key events during the presentation, such as moments of nervous movement, instances of reading from notes or slides, and episodes of eye contact with the audience.
\end{itemize}   

\begin{figure}[th]
    \centering
    \includegraphics[width=1\linewidth]{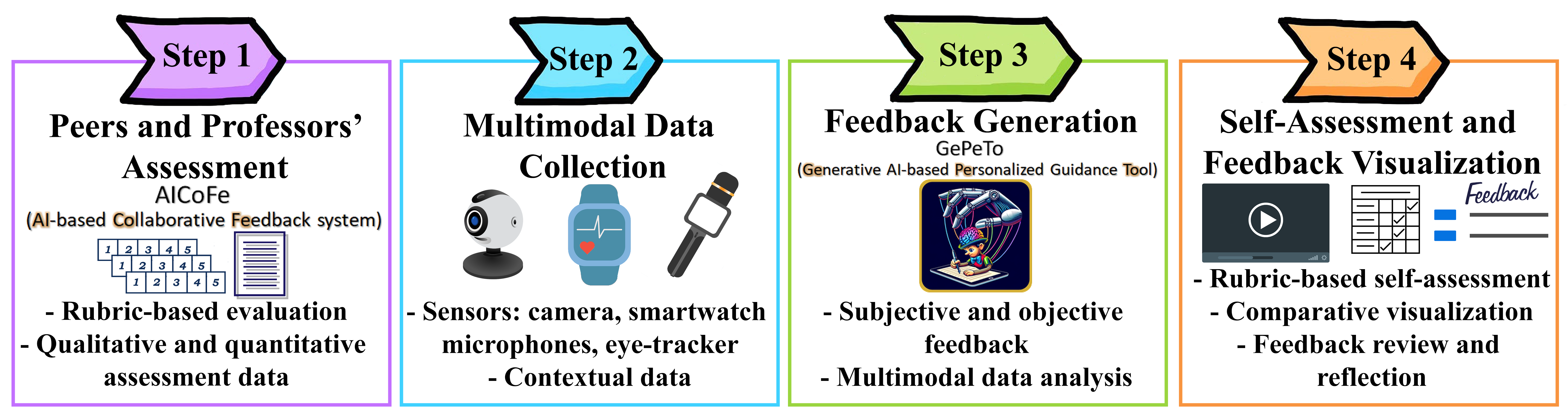}
    \caption{MOSAIC-F workflow diagram.}
    \label{fig:workflow}
\end{figure}

MOSAIC-F use a four step workflow (Figure \ref{fig:workflow}):
\begin{enumerate}
    \item Peers and Professors' Assessment (see Subsection \ref{ss:step1})
    \item Multimodal Data Collection (see Subsection \ref{ss:step2})
    \item Feedback Generation (see Subsection \ref{ss:step3})
    \item Self-Assessment and Feedback Visualization (see Subsection \ref{ss:step4})
\end{enumerate}

\subsection{Peers and Professors' Assessment}\label{ss:step1}
In this step, students’ performance is evaluated. In our case study, oral presentations are assessed by both professors and peers using a standardized rubric within the AICoFe system \cite{becerra2025aicofe}. AICoFe provides individual web-based dashboards for each evaluator, where the rubric is presented. These dashboards are accessible through any web browser.

The rubric includes different items, such as eye contact, attention capture or clarity of the opening, and each item is assessed using a 5-point Likert scale, with predefined descriptions for each level to ensure consistency among all the evaluators. Additionally, peers and professors should provide qualitative observations for each item.

\subsection{Multimodal Data Collection}\label{ss:step2}

\begin{figure}[th]
    \centering
    \includegraphics[width=0.95\linewidth]{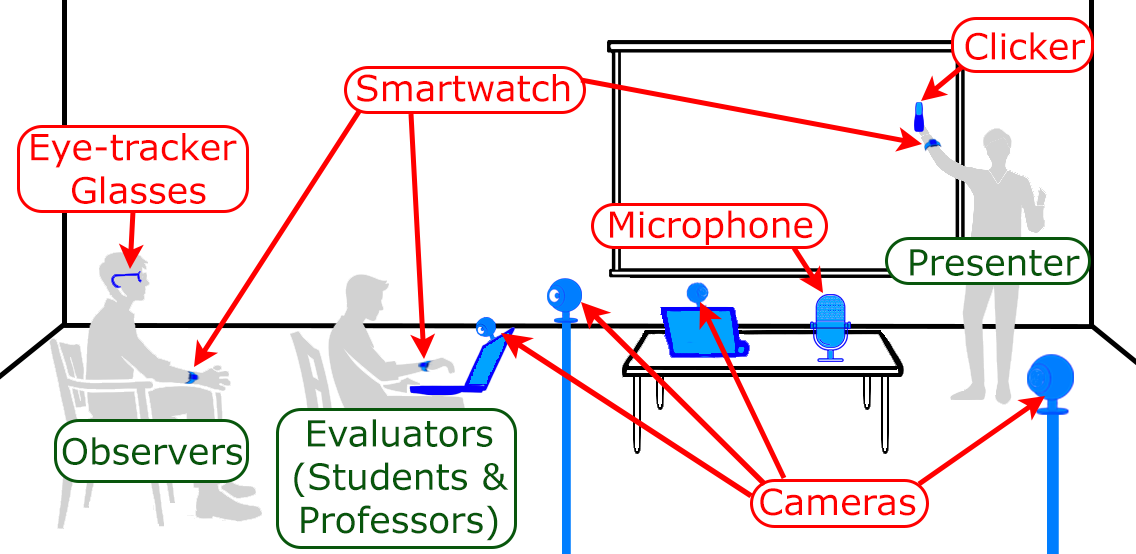}
    \caption{Acquisition setup illustrating the different roles and sensors used.}
    \label{fig:setup}
\end{figure}

MOSAIC-F uses several sensors and data as illustrated in Figure~\ref{fig:setup} to monitor students and gain data-based insights into their performance while delivering an oral presentation. In particular, the following sensors are used:
\begin{itemize}
    \item \textbf{Two Logitech C920 PRO HD Webcams:} One RGB webcam records the presenter’s performance from a front-facing view. The second webcam records the evaluators and the external observer during the oral presentation. Through the edBB platform~\cite{daza2023edbb}, both audio and video from these webcams are captured.
    \item \textbf{Integrated Webcams:} Presenters and evaluators are recorded using Microsoft Teams and the webcams integrated in their laptops.
    \item \textbf{Poly SYNC 40 Microphone:} This non-intrusive ambient microphone is placed near the student to capture high-quality audio of the oral presentation using Microsoft Teams.
    \item \textbf{Fitbit Sense Smartwatch:} The presenter, evaluators, and external observer wear a smartwatch that captures heart rate and motion data, including gyroscope, accelerometer, and device orientation.
    \item \textbf{Clicker, Mouse and Keyboard:} Presenters can use a clicker, the mouse or the keyboard to advance the slides. All interactions with these devices are stored.
    \item \textbf{Keyboard and Interactions Logs:} While evaluators assess oral presentations using AICoFe \cite{becerra2025aicofe}, all clicks and keystrokes made on the platform are recorded.
    \item \textbf{Tobii Pro Glasses 3: } An observer wears an eye-tracking device that captures data on gaze, fixations, and saccades.
    \item \textbf{Contextual Data Annotations:} An observer (research assistant) uses a custom interface to label contextual events during the presentation in real time, such as instances of nervous movement, reading behavior, and moments of eye contact. These structured annotations enrich the multimodal dataset and can be later integrated into machine learning models for behavior analysis.
\end{itemize}

\subsection{Feedback Generation}\label{ss:step3}
Once all evaluations are collected, AICoFe leverages Generative AI through GePeTo \cite{becerra2024generative} to automatically generate personalized feedback based on the quantitative and qualitative input provided by peers and professors. This feedback is structured around three core components: (1) a summary of the presenter’s strengths, (2) identification of areas for improvement, and (3) an action plan that offers concrete, targeted recommendations for enhancing oral presentation skills.

GePeTo is built on a fine-tuned version of the ChatGPT language model, specifically adapted for oral presentations. The model has been fine-tuned using feedback examples, ensuring that the generated outputs are pedagogically sound, contextually relevant, and aligned with the evaluation rubric used in AICoFe.

In addition to this human-based feedback, a data-based feedback report is generated based on MMLA, incorporating multiple analyses derived from the captured multimodal data. Section \ref{s:multimodal_analyses} presents the analyses to be included in our case study of oral presentations.

\subsection{Self-Assessment and Feedback Visualization}\label{ss:step4}

A key component of effective feedback is supporting students in reflecting on their own performance. In this step of the MOSAIC-F framework, students review video recordings of their performances as a foundation for self-assessment. This practice allows them to observe themselves from an external perspective, recognize specific behaviors, and evaluate their performance more objectively. Once the self-assessment is submitted, personalized feedback is provided along with visualizations that compare their results to those of peers and professors, as well as class averages. Finally, students are invited to reflect on the feedback received and indicate whether they agree with the assessment and find it useful for their improvement process.

In our case study, after delivering their presentation, students engage in a self-assessment process using video recordings captured from both Microsoft Teams and the frontal camera. These recordings allow them to critically review their own performance and complete the same standardized evaluation rubric used by peers and professors within the AICoFe system. Upon submitting their self-assessment, the system displays the personalized feedback. Although the feedback is automatically produced, it is subsequently reviewed by professors to ensure its accuracy, coherence, and alignment with the rubric. This human oversight helps prevent potential inconsistencies or hallucinations from the language model, ensuring that the final feedback remains pedagogically valid and trustworthy.

In addition to textual feedback, AICoFe provides interactive visualizations that allow students to compare their self-assessment results with peer and professor evaluations, as well as with class averages. This comparative analysis helps students gain deeper insight into their performance, identify discrepancies between self and external assessments, and reflect on areas for growth.

Finally, this reflection process is further supported by a final step in which students are asked to evaluate the feedback they received, indicating whether they found it accurate, useful, and relevant for improving their future presentations.

\section{Multimodal Data Analyses in the Oral Presentations Case Study}\label{s:multimodal_analyses}
In Step 3 (see Subsection \ref{ss:step3}), MOSAIC-F generates data-based feedback using multimodal and biometric data. In our case study, we are planning to conduct the following analyses:
\begin{itemize}
\item \textbf{Head Pose Analysis:} edBB and Microsoft Teams recordings will be used to estimate the Euler angles (pitch, yaw, and roll). The presenter’s head pose will then be analyzed to infer patterns of visual attention throughout the presentation. This analysis will help identify when the presenter is making eye contact with the audience, directing their gaze toward the projected slides, looking downward at the floor, or referring to written notes. Similarly, the head pose of the evaluators will also be analyzed to assess their level of attentiveness during the evaluation process. By examining the direction and stability of their gaze, it will be possible to infer when evaluators were actively focused on the presenter, when they were evaluating, and when they were distracted.

\item \textbf{Body Posture Analysis:} Once the body landmark data have been extracted, the presenter’s posture will be analyzed to identify various types of body language and the specific moments at which they occur. For instance, closed or constrained positions—such as crossed arms, hunched shoulders, or limited mobility—may convey a lack of self-confidence and negatively affect the effectiveness of the presentation. In contrast, excessive or erratic movements, including repetitive pacing or continuous shifting of weight, could signal nervousness and potentially distract the audience. On the other hand, open and stable postures will be associated with confidence, clarity, and stronger audience engagement.

\item \textbf{Audio Analysis:} The presenter’s vocal performance will be analyzed. Key audio features, including tone (pitch), modulation, fluency, and clarity, will be extracted to evaluate aspects of speech delivery. This analysis will help identify monotone speech, lack of vocal variation, and articulation issues, all of which are critical for effective oral communication.

\item \textbf{Speech Transcription and Pattern Detection:} The presenter’s speech will be automatically transcribed. This transcription will enable the detection of linguistic patterns that may affect communication quality, including the frequent use of filler words (e.g., ``um'', ``uh'', ``like'', or ``you know''), false starts, and long or frequent pauses.

\item \textbf{Heart Rate Analysis:} The presenter’s heart rate will be analyzed using data collected from the smartwatch. The signal will be processed to extract temporal patterns and identify fluctuations throughout the presentation. The analysis will focus on detecting peaks, stable periods, and abrupt changes that may correspond to key segments such as the introduction, transitions, or audience interactions. To evaluate whether heart rate significantly varies across different presentation phases, statistical tests such as paired t-tests will be applied. These tests will allow for the comparison of heart rate levels between predefined intervals (e.g., opening vs. conclusion), helping to determine whether physiological stress responses are consistently associated with specific moments of the presentation. In addition, peaks will be detected and mapped against the presentation timeline to determine whether they coincide with specific events, such as the start of the talk, slide changes, or audience questions. Heart rate data will also be analyzed for the evaluators to assess their physiological engagement during the presentation.

\item \textbf{Gaze Analysis:} Gaze data from an external observer wearing eye-tracking glasses and located in the audience will be analyzed. The recorded data will include precise information on gaze direction, fixations, eyeblinks and saccades in real-world coordinates. This information will be used to determine whether the presenter successfully captured and maintained the observer’s visual attention throughout the presentation. By mapping gaze fixations onto areas of interest (AOIs), such as the presenter’s face, the slides, or irrelevant regions, the analysis will reveal how effectively the presenter directs audience attention. Temporal patterns of gaze distribution will also be examined to identify shifts in focus, lapses in attention, or moments of distraction.

\item \textbf{Keystroke and Interaction Logs:} While evaluators assess presentations through the AICoFe system, all keystrokes and interaction events will be logged with precise timestamps. This interaction data will be used to analyze the sequence and timing of evaluation steps, providing evidence of whether the assessment process was carried out appropriately. For instance, it will be possible to detect if an evaluator rated final rubric items (such as conclusions) before the corresponding segment of the presentation occurred, indicating a possible lack of attention or premature judgment. Additionally, typing behavior and the time spent on each rubric item will be analyzed to reveal the level of engagement and depth of reflection during the evaluation. By examining these logs, the analysis will support quality assurance in the assessment process, helping to identify superficial, rushed, or inconsistent evaluations.

\item \textbf{Slides Analysis:} In addition to tracking slide transitions during the presentation, the original \texttt{.pptx} file will be analyzed to examine the structure and design of the slides. This analysis will include the extraction of key features such as the total number of slides, slide titles, use of images, presence and position of slide numbers, and the estimated text density per slide. Particular attention will be paid to the font size used in text boxes, as overly small fonts may hinder readability and negatively affect audience engagement.
\end{itemize}

\section{Conclusions and Future Work}\label{s:conclusions}
In this article, we introduced MOSAIC-F framework, a novel approach to enhancing students’ skills through the integration of Multimodal Learning Analytics, Observations, Sensors and Collaborative assessment along with Artificial Intelligence that supports the entire process. Its structured four-step process comprising peer and professor assessments, multimodal data collection, AI-driven feedback generation, and student self-assessment with feedback visualization and aims to deliver more comprehensive, personalized and actionable feedback on student performance.

By combining human-based assessments with multimodal data, such as posture, gaze, speech patterns, and physiological signals, MOSAIC-F has the potential to provide richer, more personalized feedback that goes beyond traditional evaluation methods. The inclusion of self-reflection activities and comparative visualizations further supports student engagement, metacognition, and the development of targeted improvement strategies.

We carried out a case study focused on improving oral presentation skills in a university face-to-face learning setting, applying the MOSAIC-F framework. This initial implementation allowed us to validate the framework’s feasibility and laid the groundwork for future iterations. We plan to apply MOSAIC-F to additional case studies across diverse educational contexts and competencies to further evaluate its adaptability, effectiveness, and impact on student learning.

As part of our future work, we will conduct an in-depth analysis of the multimodal data collected during the case study to assess the effectiveness and accuracy of the feedback mechanisms integrated in MOSAIC-F. Beyond this, we aim to expand the framework’s application to diverse learning scenarios and to different skill domains such as teamwork. Additionally, we plan to develop interactive tools that allow students to independently practise their skills and receive formative feedback, supporting ongoing improvement beyond formal assessment contexts.



\begin{acknowledgments}
Support by projects: Cátedra ENIA UAM-VERIDAS en IA Responsable (NextGenerationEU PRTR TSI-100927-2023-2), HumanCAIC (TED2021-131787B-I00 MICINN) and SNOLA (RED2022-134284-T).

In addition, we would like to thank the professors and students involved in the courses supported by the INNOVA project entitled ``Presentacions de Impacto'' financed by the Universidad Autónoma de Madrid.

\end{acknowledgments}

\section*{Declaration on Generative AI}
The authors used ChatGPT-4 in order to grammar and spelling check.

\bibliography{sample-ceur}

\begin{thebibliography}{37}
\expandafter\ifx\csname natexlab\endcsname\relax\def\natexlab#1{#1}\fi
\providecommand{\url}[1]{\texttt{#1}}
\providecommand{\href}[2]{#2}
\providecommand{\path}[1]{#1}
\providecommand{\DOIprefix}{doi:}
\providecommand{\ArXivprefix}{arXiv:}
\providecommand{\URLprefix}{URL: }
\providecommand{\Pubmedprefix}{pmid:}
\providecommand{\doi}[1]{\href{http://dx.doi.org/#1}{\path{#1}}}
\providecommand{\Pubmed}[1]{\href{pmid:#1}{\path{#1}}}
\providecommand{\bibinfo}[2]{#2}
\ifx\xfnm\relax \def\xfnm[#1]{\unskip,\space#1}\fi
\bibitem[{Hattie and Timperley(2007)}]{hattie2007power}
\bibinfo{author}{J.~Hattie}, \bibinfo{author}{H.~Timperley},
\newblock \bibinfo{title}{{The Power of Feedback}},
\newblock \bibinfo{journal}{Review of Educational Research} \bibinfo{volume}{77} (\bibinfo{year}{2007}) \bibinfo{pages}{81--112}.
\bibitem[{Askew and Askew(2000)}]{askew2000feedback}
\bibinfo{author}{S.~Askew}, \bibinfo{author}{S.~Askew}, \bibinfo{title}{{Feedback for Learning}}, \bibinfo{type}{Technical Report}, RoutledgeFalmer London, \bibinfo{year}{2000}.
\bibitem[{Henderson et~al.(2019)Henderson, Ryan, and Phillips}]{henderson2019challenges}
\bibinfo{author}{M.~Henderson}, \bibinfo{author}{T.~Ryan}, \bibinfo{author}{M.~Phillips},
\newblock \bibinfo{title}{{The Challenges of Feedback in Higher Education}},
\newblock \bibinfo{journal}{Assessment \& Evaluation in Higher Education}  (\bibinfo{year}{2019}).
\bibitem[{Tailab and Marsh(2020)}]{tailab2020use}
\bibinfo{author}{M.~Tailab}, \bibinfo{author}{N.~Marsh},
\newblock \bibinfo{title}{{Use of Self-Assessment of Video Recording to Raise Students' Awareness of Development of Their Oral Presentation Skills}},
\newblock \bibinfo{journal}{Higher Education Studies} \bibinfo{volume}{10} (\bibinfo{year}{2020}).
\bibitem[{Giannakos et~al.(2022)Giannakos, Spikol, Di~Mitri, Sharma, Ochoa, and Hammad}]{giannakos2022multimodal}
\bibinfo{author}{M.~Giannakos}, \bibinfo{author}{D.~Spikol}, \bibinfo{author}{D.~Di~Mitri}, \bibinfo{author}{K.~Sharma}, \bibinfo{author}{X.~Ochoa}, \bibinfo{author}{R.~Hammad}, \bibinfo{title}{{The Multimodal Learning Analytics Handbook}}, \bibinfo{publisher}{Springer}, \bibinfo{year}{2022}.
\bibitem[{Lang et~al.(2017)Lang, Siemens, Wise, and Gasevic}]{lang2017handbook}
\bibinfo{author}{C.~Lang}, \bibinfo{author}{G.~Siemens}, \bibinfo{author}{A.~Wise}, \bibinfo{author}{D.~Gasevic},
\newblock \bibinfo{title}{{Handbook of Learning Analytics}}  (\bibinfo{year}{2017}).
\bibitem[{Bar{\'o}-Sol{\'e} et~al.(2018)Bar{\'o}-Sol{\'e}, Guerrero-Roldan, Prieto-Bl{\'a}zquez, Rozeva, Marinov, Kiennert, Rocher, and Garcia-Alfaro}]{baro2018integration}
\bibinfo{author}{X.~Bar{\'o}-Sol{\'e}}, \bibinfo{author}{A.~E. Guerrero-Roldan}, \bibinfo{author}{J.~Prieto-Bl{\'a}zquez}, \bibinfo{author}{A.~Rozeva}, \bibinfo{author}{O.~Marinov}, \bibinfo{author}{C.~Kiennert}, \bibinfo{author}{P.-O. Rocher}, \bibinfo{author}{J.~Garcia-Alfaro},
\newblock \bibinfo{title}{{Integration of an Adaptive Trust-Based E-Assessment System into Virtual Learning Environments—The TeSLA Project Experience}},
\newblock \bibinfo{journal}{Internet Technology Letters} \bibinfo{volume}{1} (\bibinfo{year}{2018}) \bibinfo{pages}{e56}.
\bibitem[{Daza et~al.(2023)Daza, Morales, Tolosana, Gomez, Fierrez, and Ortega-Garcia}]{daza2023edbb}
\bibinfo{author}{R.~Daza}, \bibinfo{author}{A.~Morales}, \bibinfo{author}{R.~Tolosana}, \bibinfo{author}{L.~F. Gomez}, \bibinfo{author}{J.~Fierrez}, \bibinfo{author}{J.~Ortega-Garcia},
\newblock \bibinfo{title}{{edBB-Demo: Biometrics and Behavior Analysis for Online Educational Platforms}},
\newblock in: \bibinfo{booktitle}{{Proceedings of the AAAI Conference on Artificial Intelligence}}, volume~\bibinfo{volume}{37}, \bibinfo{year}{2023}, pp. \bibinfo{pages}{16422--16424}.
\bibitem[{Becerra et~al.(2023)Becerra, Daza, Cobos, Morales, Cukurova, and Fierrez}]{becerra2023m2lads}
\bibinfo{author}{A.~Becerra}, \bibinfo{author}{R.~Daza}, \bibinfo{author}{R.~Cobos}, \bibinfo{author}{A.~Morales}, \bibinfo{author}{M.~Cukurova}, \bibinfo{author}{J.~Fierrez},
\newblock \bibinfo{title}{{M2LADS: A System for Generating Multimodal Learning Analytics Dashboards}},
\newblock in: \bibinfo{booktitle}{{2023 IEEE 47th Annual Computers, Software, and Applications Conference (COMPSAC)}}, \bibinfo{organization}{IEEE}, \bibinfo{year}{2023}, pp. \bibinfo{pages}{1564--1569}.
\bibitem[{Becerra et~al.(2025)Becerra, Daza, Cobos, Morales, and Fierrez}]{becerra2025m2lads}
\bibinfo{author}{A.~Becerra}, \bibinfo{author}{R.~Daza}, \bibinfo{author}{R.~Cobos}, \bibinfo{author}{A.~Morales}, \bibinfo{author}{J.~Fierrez},
\newblock \bibinfo{title}{{M2LADS Demo: A System for Generating Multimodal Learning Analytics Dashboards}},
\newblock \bibinfo{journal}{arXiv preprint arXiv:2502.15363}  (\bibinfo{year}{2025}).
\bibitem[{Becerra et~al.(2023)Becerra, Daza, Cobos, Morales, and Fierrez}]{becerra2023user}
\bibinfo{author}{A.~Becerra}, \bibinfo{author}{R.~Daza}, \bibinfo{author}{R.~Cobos}, \bibinfo{author}{A.~Morales}, \bibinfo{author}{J.~Fierrez},
\newblock \bibinfo{title}{{User Experience Study Using a System for Generating Multimodal Learning Analytics Dashboards}},
\newblock in: \bibinfo{booktitle}{{Proceedings of the XXIII International Conference on Human Computer Interaction}}, \bibinfo{year}{2023}, pp. \bibinfo{pages}{1--2}.
\bibitem[{Becerra et~al.(2024)Becerra, Irigoyen, Daza, Cobos, Morales, Fierrez, and Cukurova}]{becerra2024biometrics}
\bibinfo{author}{A.~Becerra}, \bibinfo{author}{J.~Irigoyen}, \bibinfo{author}{R.~Daza}, \bibinfo{author}{R.~Cobos}, \bibinfo{author}{A.~Morales}, \bibinfo{author}{J.~Fierrez}, \bibinfo{author}{M.~Cukurova},
\newblock \bibinfo{title}{{Biometrics and Behavioral Modelling for Detecting Distractions in Online Learning}},
\newblock in: \bibinfo{booktitle}{{Proc. Simposio Internacional de Inform\'{a}tica Educativa (SIIE), VII Congreso Espa\~{n}ol de Inform\'{a}tica}}, \bibinfo{year}{2024}.
\bibitem[{Daza et~al.(2024)Daza, Becerra, Cobos, Fierrez, and Morales}]{daza2024improve}
\bibinfo{author}{R.~Daza}, \bibinfo{author}{A.~Becerra}, \bibinfo{author}{R.~Cobos}, \bibinfo{author}{J.~Fierrez}, \bibinfo{author}{A.~Morales},
\newblock \bibinfo{title}{{IMPROVE: Impact of Mobile Phones on Remote Online Virtual Education}},
\newblock \bibinfo{journal}{arXiv preprint arXiv:2412.14195}  (\bibinfo{year}{2024}).
\bibitem[{Navarro et~al.(2024)Navarro, Becerra, Daza, Cobos, Morales, and Fierrez}]{navarro2024vaad}
\bibinfo{author}{M.~Navarro}, \bibinfo{author}{A.~Becerra}, \bibinfo{author}{R.~Daza}, \bibinfo{author}{R.~Cobos}, \bibinfo{author}{A.~Morales}, \bibinfo{author}{J.~Fierrez},
\newblock \bibinfo{title}{{VAAD: Visual Attention Analysis Dashboard Applied to E-Learning}},
\newblock in: \bibinfo{booktitle}{{2024 International Symposium on Computers in Education (SIIE)}}, \bibinfo{organization}{IEEE}, \bibinfo{year}{2024}, pp. \bibinfo{pages}{1--6}.
\bibitem[{Spikol et~al.(2018)Spikol, Ruffaldi, Dabisias, and Cukurova}]{spikol2018supervised}
\bibinfo{author}{D.~Spikol}, \bibinfo{author}{E.~Ruffaldi}, \bibinfo{author}{G.~Dabisias}, \bibinfo{author}{M.~Cukurova},
\newblock \bibinfo{title}{{Supervised Machine Learning in Multimodal Learning Analytics for Estimating Success in Project-Based Learning}},
\newblock \bibinfo{journal}{Journal of Computer Assisted Learning} \bibinfo{volume}{34} (\bibinfo{year}{2018}) \bibinfo{pages}{366--377}.
\bibitem[{Garc\'{\i}a et~al.(2024)Garc\'{\i}a, C\'{a}novas, and Clemente}]{garcia2024exploring}
\bibinfo{author}{F.~P. Garc\'{\i}a}, \bibinfo{author}{O.~C\'{a}novas}, \bibinfo{author}{F.~J.~G. Clemente},
\newblock \bibinfo{title}{{Exploring AI Techniques for Generalizable Teaching Practice Identification}},
\newblock \bibinfo{journal}{IEEE Access}  (\bibinfo{year}{2024}).
\bibitem[{Bosch et~al.(2014)Bosch, Chen, and D'Mello}]{bosch2014s}
\bibinfo{author}{N.~Bosch}, \bibinfo{author}{Y.~Chen}, \bibinfo{author}{S.~D'Mello},
\newblock \bibinfo{title}{{It's Written on Your Face: Detecting Affective States from Facial Expressions While Learning Computer Programming}},
\newblock in: \bibinfo{booktitle}{{Intelligent Tutoring Systems: 12th International Conference, ITS 2014, Honolulu, HI, USA, June 5-9, 2014. Proceedings 12}}, \bibinfo{organization}{Springer}, \bibinfo{year}{2014}, pp. \bibinfo{pages}{39--44}.
\bibitem[{Ekin et~al.(2025)Ekin, Cantekin, Polat, and Hopcan}]{ekin2025artificial}
\bibinfo{author}{C.~C. Ekin}, \bibinfo{author}{O.~F. Cantekin}, \bibinfo{author}{E.~Polat}, \bibinfo{author}{S.~Hopcan},
\newblock \bibinfo{title}{{Artificial Intelligence in Education: A Text Mining-Based Review of the Past 56 Years}},
\newblock \bibinfo{journal}{Education and Information Technologies}  (\bibinfo{year}{2025}) \bibinfo{pages}{1--43}.
\bibitem[{Kim et~al.(2024)Kim, Kim, Lee, Yoon, Myung, Yoo, Lim, Han, Kim, Ahn et~al.}]{kim2024llm}
\bibinfo{author}{M.~Kim}, \bibinfo{author}{S.~Kim}, \bibinfo{author}{S.~Lee}, \bibinfo{author}{Y.~Yoon}, \bibinfo{author}{J.~Myung}, \bibinfo{author}{H.~Yoo}, \bibinfo{author}{H.~Lim}, \bibinfo{author}{J.~Han}, \bibinfo{author}{Y.~Kim}, \bibinfo{author}{S.-Y. Ahn}, et~al.,
\newblock \bibinfo{title}{{LLM-Driven Learning Analytics Dashboard for Teachers in EFL Writing Education}},
\newblock \bibinfo{journal}{arXiv preprint arXiv:2410.15025}  (\bibinfo{year}{2024}).
\bibitem[{Schneider et~al.(2017)Schneider, B{\"o}rner, Van~Rosmalen, and Specht}]{schneider2017presentation}
\bibinfo{author}{J.~Schneider}, \bibinfo{author}{D.~B{\"o}rner}, \bibinfo{author}{P.~Van~Rosmalen}, \bibinfo{author}{M.~Specht},
\newblock \bibinfo{title}{{Presentation Trainer: What Experts and Computers Can Tell About Your Nonverbal Communication}},
\newblock \bibinfo{journal}{Journal of Computer Assisted Learning} \bibinfo{volume}{33} (\bibinfo{year}{2017}) \bibinfo{pages}{164--177}.
\bibitem[{Di~Mitri et~al.(2025)Di~Mitri, Schneider, Mouhammad, Hummel, Alomari, Ali, Masum, Arif, Rose, and Klemke}]{di2025presentable}
\bibinfo{author}{D.~Di~Mitri}, \bibinfo{author}{J.~Schneider}, \bibinfo{author}{N.~Mouhammad}, \bibinfo{author}{S.~Hummel}, \bibinfo{author}{M.~Alomari}, \bibinfo{author}{M.~A. Ali}, \bibinfo{author}{M.~H.~R. Masum}, \bibinfo{author}{H.~Arif}, \bibinfo{author}{M.~Rose}, \bibinfo{author}{R.~Klemke},
\newblock \bibinfo{title}{{Enhance Your Presentation Skills with Presentable}},
\newblock in: \bibinfo{booktitle}{{Proceedings of the 15th International Conference on Learning Analytics \& Knowledge (LAK 2025), Demo Track}}, \bibinfo{address}{Dublin, Ireland}, \bibinfo{year}{2025}. \bibinfo{note}{March 3--7}.
\bibitem[{Ochoa et~al.(2018)Ochoa, Dom{\'\i}nguez, Guam{\'a}n, Maya, Falcones, and Castells}]{ochoa2018rap}
\bibinfo{author}{X.~Ochoa}, \bibinfo{author}{F.~Dom{\'\i}nguez}, \bibinfo{author}{B.~Guam{\'a}n}, \bibinfo{author}{R.~Maya}, \bibinfo{author}{G.~Falcones}, \bibinfo{author}{J.~Castells},
\newblock \bibinfo{title}{{The RAP System: Automatic Feedback of Oral Presentation Skills Using Multimodal Analysis and Low-Cost Sensors}},
\newblock in: \bibinfo{booktitle}{{Proceedings of the 8th International Conference on Learning Analytics and Knowledge}}, \bibinfo{year}{2018}, pp. \bibinfo{pages}{360--364}.
\bibitem[{Ochoa and Zhao(2024)}]{ochoa2024openopaf}
\bibinfo{author}{X.~Ochoa}, \bibinfo{author}{H.~Zhao},
\newblock \bibinfo{title}{{OpenOPAF: An Open-Source Multimodal System for Automated Feedback for Oral Presentations}},
\newblock \bibinfo{journal}{Journal of Learning Analytics} \bibinfo{volume}{11} (\bibinfo{year}{2024}) \bibinfo{pages}{224--248}.
\bibitem[{Daza et~al.(2025)Daza, Shengkai, Morales, Fierrez, and Nagao}]{daza2025smartevr}
\bibinfo{author}{R.~Daza}, \bibinfo{author}{L.~Shengkai}, \bibinfo{author}{A.~Morales}, \bibinfo{author}{J.~Fierrez}, \bibinfo{author}{K.~Nagao},
\newblock \bibinfo{title}{{{SMARTe-VR: Student Monitoring and Adaptive Response Technology for e-learning in Virtual Reality}}},
\newblock in: \bibinfo{booktitle}{Proc. AAAI Workshop on Artificial Intelligence for Education}, \bibinfo{year}{2025}.
\bibitem[{Yokoyama and Nagao(2021)}]{yokoyama2021vr}
\bibinfo{author}{Y.~Yokoyama}, \bibinfo{author}{K.~Nagao},
\newblock \bibinfo{title}{{VR Presentation Training System Using Machine Learning Techniques for Automatic Evaluation}},
\newblock \bibinfo{journal}{International Journal of Virtual and Augmented Reality (IJVAR)}  (\bibinfo{year}{2021}).
\bibitem[{VanLehn(2011)}]{vanlehn2011relative}
\bibinfo{author}{K.~VanLehn},
\newblock \bibinfo{title}{{The Relative Effectiveness of Human Tutoring, Intelligent Tutoring Systems, and Other Tutoring Systems}},
\newblock \bibinfo{journal}{Educational Psychologist} \bibinfo{volume}{46} (\bibinfo{year}{2011}) \bibinfo{pages}{197--221}.
\bibitem[{Zhang and Litman(2019)}]{zhang2019co}
\bibinfo{author}{H.~Zhang}, \bibinfo{author}{D.~Litman},
\newblock \bibinfo{title}{{Co-Attention Based Neural Network for Source-Dependent Essay Scoring}},
\newblock \bibinfo{journal}{arXiv preprint arXiv:1908.01993}  (\bibinfo{year}{2019}).
\bibitem[{R\"{u}dian et~al.(2025)R\"{u}dian, Podelo, Ku\v{z}\'{\i}lek, and Pinkwart}]{rudian2025feedback}
\bibinfo{author}{S.~R\"{u}dian}, \bibinfo{author}{J.~Podelo}, \bibinfo{author}{J.~Ku\v{z}\'{\i}lek}, \bibinfo{author}{N.~Pinkwart},
\newblock \bibinfo{title}{{Feedback on Feedback: Student’s Perceptions for Feedback from Teachers and Few-Shot LLMs}},
\newblock in: \bibinfo{booktitle}{{Proceedings of the 15th International Learning Analytics and Knowledge Conference}}, \bibinfo{year}{2025}, pp. \bibinfo{pages}{82--92}.
\bibitem[{Nazaretsky et~al.(2024)Nazaretsky, Mejia-Domenzain, Swamy, Frej, and K\"{a}ser}]{nazaretsky2024ai}
\bibinfo{author}{T.~Nazaretsky}, \bibinfo{author}{P.~Mejia-Domenzain}, \bibinfo{author}{V.~Swamy}, \bibinfo{author}{J.~Frej}, \bibinfo{author}{T.~K\"{a}ser},
\newblock \bibinfo{title}{{AI or Human? Evaluating Student Feedback Perceptions in Higher Education}},
\newblock in: \bibinfo{booktitle}{{European Conference on Technology Enhanced Learning}}, \bibinfo{organization}{Springer}, \bibinfo{year}{2024}, pp. \bibinfo{pages}{284--298}.
\bibitem[{Nazaretsky et~al.(2025)Nazaretsky, Mejia-Domenzain, Swamy, Frej, and K\"{a}ser}]{nazaretsky2025critical}
\bibinfo{author}{T.~Nazaretsky}, \bibinfo{author}{P.~Mejia-Domenzain}, \bibinfo{author}{V.~Swamy}, \bibinfo{author}{J.~Frej}, \bibinfo{author}{T.~K\"{a}ser},
\newblock \bibinfo{title}{{The Critical Role of Trust in Adopting AI-Powered Educational Technology for Learning: An Instrument for Measuring Student Perceptions}},
\newblock \bibinfo{journal}{Computers and Education: Artificial Intelligence}  (\bibinfo{year}{2025}) \bibinfo{pages}{100368}.
\bibitem[{Steiss et~al.(2024)Steiss, Tate, Graham, Cruz, Hebert, Wang, Moon, Tseng, Warschauer, and Olson}]{steiss2024comparing}
\bibinfo{author}{J.~Steiss}, \bibinfo{author}{T.~Tate}, \bibinfo{author}{S.~Graham}, \bibinfo{author}{J.~Cruz}, \bibinfo{author}{M.~Hebert}, \bibinfo{author}{J.~Wang}, \bibinfo{author}{Y.~Moon}, \bibinfo{author}{W.~Tseng}, \bibinfo{author}{M.~Warschauer}, \bibinfo{author}{C.~B. Olson},
\newblock \bibinfo{title}{{Comparing the Quality of Human and ChatGPT Feedback of Students’ Writing}},
\newblock \bibinfo{journal}{Learning and Instruction} \bibinfo{volume}{91} (\bibinfo{year}{2024}) \bibinfo{pages}{101894}.
\bibitem[{Wan and Chen(2024)}]{wan2024exploring}
\bibinfo{author}{T.~Wan}, \bibinfo{author}{Z.~Chen},
\newblock \bibinfo{title}{{Exploring Generative AI Assisted Feedback Writing for Students’ Written Responses to a Physics Conceptual Question with Prompt Engineering and Few-Shot Learning}},
\newblock \bibinfo{journal}{Physical Review Physics Education Research} \bibinfo{volume}{20} (\bibinfo{year}{2024}) \bibinfo{pages}{010152}.
\bibitem[{Kloos et~al.(2024)Kloos, Alario-Hoyos, Est\'{e}vez-Ayres, Callejo-Pinardo, Hombrados-Herrera, Mu\~{n}oz Merino, Moreno-Marcos, Mu\~{n}oz Organero, and Ib\'{a}\~{n}ez}]{kloos2024can}
\bibinfo{author}{C.~D. Kloos}, \bibinfo{author}{C.~Alario-Hoyos}, \bibinfo{author}{I.~Est\'{e}vez-Ayres}, \bibinfo{author}{P.~Callejo-Pinardo}, \bibinfo{author}{M.~A. Hombrados-Herrera}, \bibinfo{author}{P.~J. Mu\~{n}oz Merino}, \bibinfo{author}{P.~M. Moreno-Marcos}, \bibinfo{author}{M.~Mu\~{n}oz Organero}, \bibinfo{author}{M.~B. Ib\'{a}\~{n}ez},
\newblock \bibinfo{title}{{How Can Generative AI Support Education?}},
\newblock in: \bibinfo{booktitle}{{2024 IEEE Global Engineering Education Conference (EDUCON)}}, \bibinfo{organization}{IEEE}, \bibinfo{year}{2024}, pp. \bibinfo{pages}{1--7}.
\bibitem[{Ogata et~al.(2024)Ogata, Liang, Toyokawa, Hsu, Nakamura, Yamauchi, Flanagan, Dai, Takami, Horikoshi et~al.}]{ogata2024co}
\bibinfo{author}{H.~Ogata}, \bibinfo{author}{C.~Liang}, \bibinfo{author}{Y.~Toyokawa}, \bibinfo{author}{C.-Y. Hsu}, \bibinfo{author}{K.~Nakamura}, \bibinfo{author}{T.~Yamauchi}, \bibinfo{author}{B.~Flanagan}, \bibinfo{author}{Y.~Dai}, \bibinfo{author}{K.~Takami}, \bibinfo{author}{I.~Horikoshi}, et~al.,
\newblock \bibinfo{title}{{Co-Designing Data-Driven Educational Technology and Practice: Reflections from the Japanese Context}},
\newblock \bibinfo{journal}{Technology, Knowledge and Learning} \bibinfo{volume}{29} (\bibinfo{year}{2024}) \bibinfo{pages}{1711--1732}.
\bibitem[{Topali et~al.(2023)Topali, Ortega-Arranz, Dimitriadis, Villagr{\'a}-Sobrino, Mart{\'\i}nez-Mon{\'e}s, and Asensio-P{\'e}rez}]{topali2023unlock}
\bibinfo{author}{P.~Topali}, \bibinfo{author}{A.~Ortega-Arranz}, \bibinfo{author}{Y.~Dimitriadis}, \bibinfo{author}{S.~Villagr{\'a}-Sobrino}, \bibinfo{author}{A.~Mart{\'\i}nez-Mon{\'e}s}, \bibinfo{author}{J.~I. Asensio-P{\'e}rez},
\newblock \bibinfo{title}{{Unlock the feedback potential: Scaling effective teacher-led interventions in massive educational contexts}},
\newblock in: \bibinfo{booktitle}{Innovating Assessment and Feedback Design in Teacher Education}, \bibinfo{publisher}{Routledge}, \bibinfo{year}{2023}, pp. \bibinfo{pages}{1--19}.
\bibitem[{Becerra and Cobos(2025)}]{becerra2025aicofe}
\bibinfo{author}{A.~Becerra}, \bibinfo{author}{R.~Cobos},
\newblock \bibinfo{title}{{Enhancing the Professional Development of Engineering Students Through an AI-Based Collaborative Feedback System}},
\newblock in: \bibinfo{booktitle}{{2025 IEEE Global Engineering Education Conference (EDUCON)}}, \bibinfo{organization}{IEEE}, \bibinfo{year}{2025}, pp. \bibinfo{pages}{1--9}.
\bibitem[{Becerra et~al.(2024)Becerra, Mohseni, Sanz, and Cobos}]{becerra2024generative}
\bibinfo{author}{A.~Becerra}, \bibinfo{author}{Z.~Mohseni}, \bibinfo{author}{J.~Sanz}, \bibinfo{author}{R.~Cobos},
\newblock \bibinfo{title}{{A Generative AI-Based Personalized Guidance Tool for Enhancing the Feedback to MOOC Learners}},
\newblock in: \bibinfo{booktitle}{{2024 IEEE Global Engineering Education Conference (EDUCON)}}, \bibinfo{organization}{IEEE}, \bibinfo{year}{2024}, pp. \bibinfo{pages}{1--8}.

\end{thebibliography}


\end{document}